\input{aipcheck}

\documentclass[
    ,final            
  ]
  {aipproc}

\layoutstyle{6x9}

\begin{document}

\newcommand{\nablav}{\mbox{\boldmath$\nabla$}}
\newcommand{\de}{\mbox{d}}
\newcommand{\msun}{\ensuremath{{{M}}_{\scriptscriptstyle \odot}}}
\newcommand{\mhalo}{\ensuremath{M_h}}
\newcommand{\msunyr}{\ensuremath{{{M}}_{\scriptscriptstyle \odot}}/\mbox{yr}}
\newcommand{\mdot}{\dot{M}}
\newcommand{\msigma}{\ensuremath{M_\mathrm{BH}}-{\sigma}}
\newcommand{\mvc}{\ensuremath{M_\mathrm{BH}}--\ensuremath{V_c}}
\newcommand{\kms}      {\ensuremath{~\mathrm{km~s^{-1}}}} 

\newcommand{\beq}{\begin{equation}}
\newcommand{\eeq}{\end{equation}}
\def\Omm{{\Omega_m}}
\def\Ommz{{\Omega_m^{\,z}}}
\def\Omr{{\Omega_r}}
\def\Omk{{\Omega_k}}
\def\Oml{{\Omega_{\Lambda}}}

\title{The mass assembly history of black holes in the Universe}


\author{Priyamvada Natarajan}{address={Department of Astronomy, Yale University, 260 Whitney Avenue, New Haven CT 06511}}

\begin{abstract}
We track the growth and evolution of high redshift seed black holes 
over cosmic time. This population of massive, initial black hole seeds form at 
these early epochs from the direct collapse of pre-galactic gas discs. Populating 
dark matter halos with seeds formed in this fashion, we follow their mass assembly 
history to the present time using a Monte-Carlo merger tree approach. Using this 
formalism, we predict the black hole mass function at the present time; the integrated 
mass density of black holes in the Universe; the luminosity function of accreting black 
holes as a function of redshift and the scatter in observed, local $M_{\rm bh}-\sigma$ relation. 
Signatures of these massive seed models appear predominantly at the low
mass end of the present day black hole mass function. In fact, our prediction of the shape of
the $M_{\rm bh} - \sigma$ relation at the low mass end and increased
scatter has recently been corroborated by observations. These models predict 
that low surface brightness, bulge-less  galaxies with large discs are least likely to 
be sites for the formation of massive seed black holes at high redshifts. The efficiency 
of seed formation at high redshifts also has a direct influence on the black hole
occupation fraction in galaxies at $z=0$. This effect is more pronounced for low mass 
galaxies today as we predict the existence of a population of low mass galaxies that do 
not host nuclear black holes.  This is the key discriminant between the models studied 
here and the Population-III remnant seed model.   
\end{abstract}

\maketitle


\section{Introduction}

The demography of nearby galaxies suggests that most of them
harbor a quiescent super-massive black hole (SMBH) in their nucleus 
at the present time and that the mass of the hosted SMBH correlates 
with the properties of the host bulge. Observational evidence points to the 
existence of a strong correlation between the mass of the central SMBH 
and the velocity dispersion of the host spheroid (Tremaine et al. 2002;
Ferrarese \& Merritt 2000, Gebhardt et al. 2003; Marconi \& Hunt 2003;
H\'aring \& Rix 2004; G\"ultekin et al. 2009; Kormendy \& Bender 2011) and 
possibly the host halo (Ferrarese 2002; Volonteri, Natarajan \& G\"ultekin 2011) in 
nearby galaxies. These correlations plotted in Fig.~1 are strongly  suggestive of 
co-eval growth of the SMBH and the stellar component of the host galaxy. This 
 likely occurs via regulation of the gas supply in the galactic nucleus from the 
 earliest times (Haehnelt, Natarajan, Rees 1998; Silk \& Rees 1999; Kauffmann \& Haehnelt 2000; 
Fabian 2002; King 2003; Thompson, Quataert \& Murray 2005; 
Natarajan \& Treister 2009; Treister et al. 2011). 

\begin{figure}
\includegraphics[width=14cm]{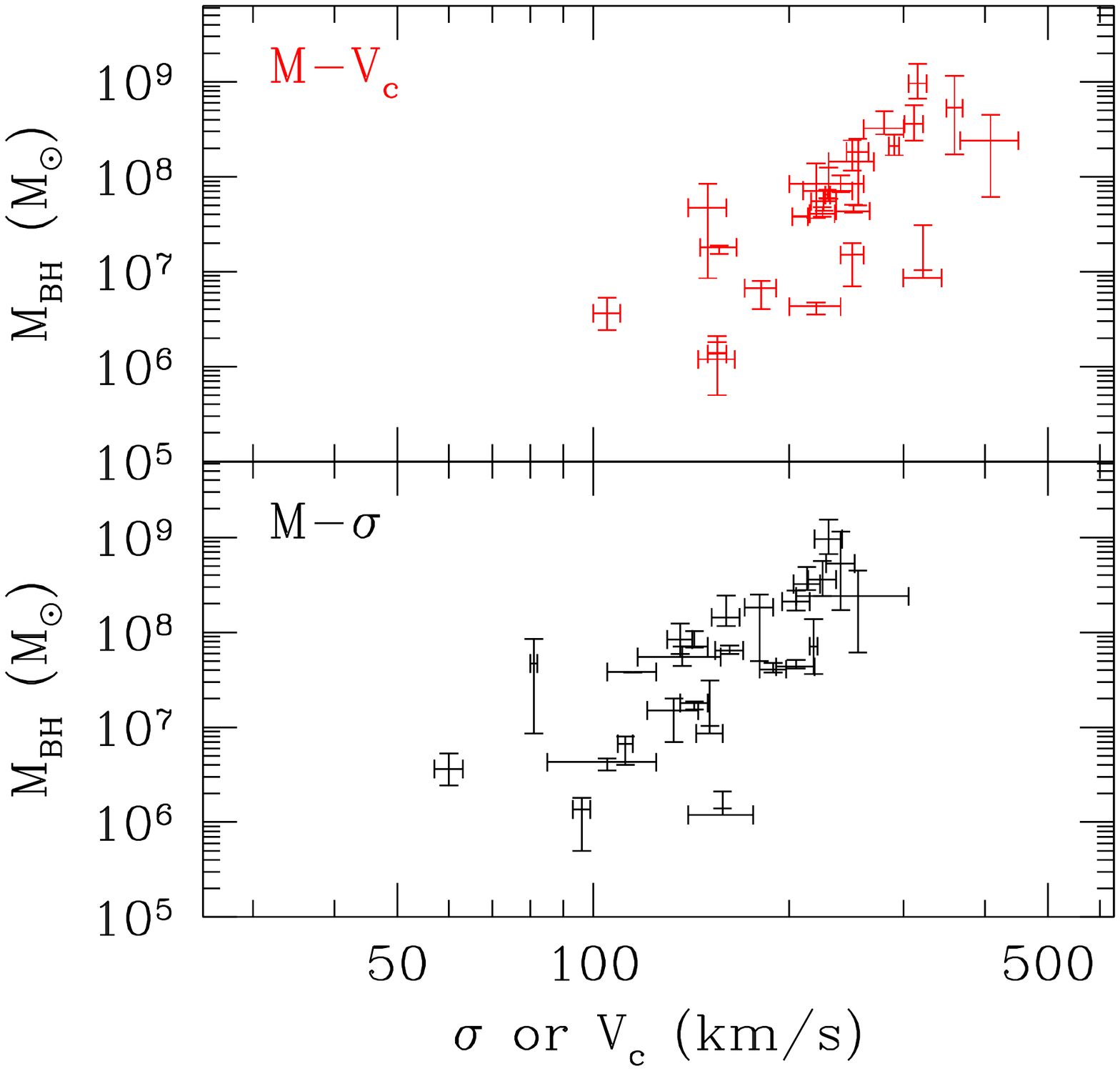}
\caption{Top panel: $M_{\rm BH}-V_c$ relation for the 25 galaxies from recently published  
Table 1of Kormendy et al. (2011). Bottom panel: The $M_{\rm BH}-\sigma$ relation from 
Kormendy's table. We note that M33 is omitted from this data set.}
\label{fig1}
\end{figure}

Black hole growth is believed to be powered by gas accretion (Lynden-Bell 1969) and actively 
accreting black holes are detected as optically  bright quasars. These optically bright quasars 
appear to exist out to the highest redshifts probed at the present time. Therefore, the mass build-up 
of SMBHs is likely to have commenced at extremely high redshifts ($z > 10$). In fact, optical
quasars have now been detected at $z > 6$ (e.g., Fan et al. 2004; 2006) in 
the Sloan Digital Sky Survey (SDSS).  Hosts of high redshift quasars 
 are often strong sources of dust emission (Omont et al. 2001; Cox et al. 2002; Carilli et al. 2002; Walter et al. 2003),
 suggesting that quasars were already in place in massive galaxies at a time
 when galaxies were undergoing vigorous star formation. Furthermore, the growth
 spurts of SMBHs are also seen in the X-ray waveband. The integrated
 emission from these X-ray quasars generates the cosmic X-ray
 background (XRB), and its spectrum suggests that most black-hole
 growth is obscured in optical wavelengths (Fabian \& Iwasawa 1999; Mushotzky 
 et al. 2000; Hasinger et al. 2001; Barger et al. 2003, 2005; Worsley et al. 2005; Treister \& Urry 2006). 
 There exist examples of obscured black-hole growth in the form of
 `Type-2' quasars, but their detected numbers are fewer than expected
 from models of the XRB. However, there is tantalizing evidence from
 infra-red (IR) studies that dust-obscured accretion is ubiquitous (Martinez-Sansigre 
 et al. 2005; Treister et al. 2006). Current work suggests that while SMBHs might
spend most of their lifetime in an optically dim phase, the bulk of their
mass growth occurs in the short-lived quasar stages (Treister et al. 2010). The assembly 
of BH mass in the Universe has been tracked using optical quasar activity. The current 
phenomenological approach to understanding the assembly of SMBHs involves optical data from both
high and low redshifts. These data are used as a starting point to
construct a consistent picture that fits within the larger framework
of the growth and evolution of structure in the Universe (Haehnelt et al. 1998; Haiman \& Loeb
1998; Kauffmann \& Haehnelt 2000, 2002; Wyithe \& Loeb 2002; Volonteri et al. 2003; Di Matteo
et al. 2003; Volonteri, Lodato \& Natarajan 2008). Optically bright quasars powered by accretion onto black 
holes are now detected out to redshifts of $z > 6$ when the Universe was barely 7\% of its current age (Fan et al. 2004; 
2006).  The luminosities of these high redshift quasars imply black hole masses $M_{\rm
  BH} > 10^9\,M_{\odot}$.

  Models that describe the growth and accretion history of supermassive black holes typically use 
  as initial seeds the remnants derived from Population-III stars (e.g. Haiman \& Loeb 1998; Haehnelt, Natarajan \& Rees 1998).  
Assembling these large black hole masses inferred at $z ~ 6$, starting from remnants of the first 
generation of metal free stars has been a challenge for models.  Copious growth episodes
are required at early times to do so. Some suggestions to accomplish this rapid growth invoke 
super-Eddington accretion rates for brief periods of time (Volonteri \& Rees 2005). Alternatively, it has been
suggested that the formation of more massive seeds ab-initio through direct  
collapse of self-gravitating pre-galactic disks might offer a new channel. Such a model has been 
proposed by  Lodato \& Natarajan 2006 [LN06]. This scenario alleviates the problem of building up
supermassive black hole masses to the required values by $z = 6$ without Super-Eddington accretion.

Current modeling is grounded in the framework of the standard paradigm that involves 
the growth of structure via gravitational amplification of small perturbations in a CDM 
Universe---a model that has independent validation, most recently from {\it Wilkinson Microwave
Anisotropy Probe} (WMAP) measurements of the anisotropies in the
cosmic microwave background (Dunkley et al. 2009). Structure
formation is tracked over cosmic time by keeping a census of the number
of collapsed dark matter halos of a given mass that form; these
provide the sites for harboring black holes. The computation of the
mass function of dark matter halos is done using either the
Press-Schechter (Press \& Schechter 1974) or the extended Press-Schechter theory 
(Lacey \& Cole 1993), or Monte-Carlo realizations of merger trees (Kauffmann \& Haehnelt 2000;
Volonteri et al. 2003; Bromley et al. 2004) or, in some cases,
directly from cosmological N-body simulations (Di Matteo et al. 2003, 2005).

In particular (Volonteri et al. 2003) have presented a detailed
merger-tree based scenario to trace the growth of black holes from the
earliest epochs to the present day. Monte-Carlo merger trees are
created for present day halos and propagated back in time to a
redshift of $\sim$ 20. With the merging history thus determined, the
initial halos at $z \sim 20$ are then populated with seed black holes
which are assumed to be remnants of the first stars that form in the
Universe. The masses of these so-called Population-III stars are not
accurately known, however numerical simulations by various groups
(Abel et al 2000; Bromm et al. 2002) suggest that they are skewed to masses of
the order of a few hundred solar masses. Seeded with the end products
of this first population, the merger sequence is followed and black
holes are assumed to grow with every major merging episode.  An
accretion episode is assumed to occur as a consequence of every major merger
event\footnote{Major mergers are 
defined as mergers between two dark matter
halos with mass ratio between 1 and 10.}. Following the growth and mass assembly of these black holes, it
is required that the model is in consonance with the observed local
$M_{\rm BH} - \sigma$ relation. The luminosity function of quasars is
predicted by these models and can be compared to
observations. It is found that not every halo at high
redshift needs to be populated with a black hole seed in order to
satisfy the observational constraints at $z = 0$. These models do not
automatically reproduce the required abundance of supermassive black
holes inferred to power the observed $z > 6$ SDSS quasars. In order to
match the observation and produce SMBHs roughly 1 Gyr after the Big
Bang, it is required that black holes undergo brief, but extremely
strong growth episodes during which the accretion rate onto them is
well in excess of the Eddington rate (Volonteri \& Rees 2005; Begelman et al. 2006).
It is the discovery and existence of this population of SMBHs powering quasars at $z > 6$ that
has prompted work on alternate channels to explain their mass
build-up. In order to alleviate the problem of explaining the existence of SMBHs
in place by $z \sim 6$, roughly 1 Gyr after the Big Bang, 
we examined the possibility of using a well motivated high
redshift massive seed black hole initial mass function. We 
populate early dark matter halos with massive black hole seeds
predicted in a model proposed by Lodato \& Natarajan (2006 (LN06 hereafter); 2007). 
This model predicts a mass function for black holes that results from the direct
collapse of pre-galactic gas discs. The implications of the
use of this massive seed mass function versus that of the Population-III
remnants, in particular, for predictions at $z = 0$ has been studied 
in detail.

In this review, we present the results of this exercise and outline the evolution of 
high redshift seed black hole masses to late times and their observational signatures.  
Populating dark matter halos with initial seeds formed in this way, we follow the mass assembly of
these black holes to the present time using the Monte-Carlo merger
tree approach. Using this machinery we predict the black hole mass function at
high redshifts and at the present time; the integrated mass density of
black holes and the luminosity function of accreting black holes as a
function of redshift. These predictions are made for a set of three
seed models with varying black hole formation efficiency. 

\section{The formation of seed black holes at high redshift}

We focus on the main features of massive seed models here.  Most aspects of the 
evolution and assembly history of this scenario have been explored in detail in 
Volonteri \& Natarajan (2009) and Volonteri, Lodato \& Natarajan (2008). In these models, at 
early times the properties of the  assembling SMBH seeds are more tightly coupled to properties 
of the dark matter halo as their growth is driven by the merger history of halos. However, at later times, when 
the merger rates are low, the final mass of the SMBH is likely more tightly coupled to the small scale 
local baryonic distribution.  The relevant host dark matter halo property at high redshifts in this 
picture is the spin or the angular momentum content. 

In a physically motivated model for the formation of heavy SMBH seeds as described 
in LN06, there is a limited range of dark matter halo spins and halo masses that are viable 
sites for their formation. In this picture, massive seeds with $M\approx 10^5-10^6M_{\odot}$ 
can form at high redshift ($z>15$), when the
intergalactic medium has not been significantly enriched by metals
(Koushiappas et al. 2004; Begelman, Volonteri \& Rees 2006;  LN06; 
Lodato \& Natarajan 2007).  As derived in LN06, it is the
development of non-axisymmetric spiral structures that drives mass infall
and accumulation in a pre-galactic disc with primordial
composition. 

The main parameter characterizing a dark matter halo that is relevant to
the fate of the gas is its spin parameter $\lambda$ ($\equiv J_h E_h^{1/2}/
GM_h^{5/2}$, where $J_h$ is the total angular momentum and $E_h$ is
the binding energy). The distribution of spin parameters for dark matter halos 
measured in numerical simulations is well fit by a lognormal distribution in
$\lambda$, with mean $\bar \lambda=0.05$ and
standard deviation $\sigma_\lambda=0.5$:
\begin{equation} 
\label{plambda}
p(\lambda) \,d\lambda
={1\over \sqrt{2\pi} \sigma_\lambda}
\exp \left[-{\ln ^2 (\lambda/{\bar \lambda})
\over 2 \sigma_\lambda^2}\right] {d\lambda \over \lambda},
\end{equation}
This function has been shown to provide a good fit to the N-body results of
several investigations at low redshift (e.g., Warren et al. 1992; Cole et al. 1996; Bullock et al. 2001;
van den Bosch et al. 2002) as well as these high redshifts (Davis \& Natarajan 2009; 2010). If the virial 
temperature of the halo $T_{\rm vir}>T_{\rm gas}$, the gas collapses and forms a rotationally supported 
disc. For low values of the spin parameter $\lambda$ the resulting disc can be compact and
dense and is subject to gravitational instabilities. This occurs when
the stability parameter $Q$ defined below approaches unity:
\begin{equation}
Q=\frac{c_{\rm s}\kappa}{\pi G \Sigma}=\sqrt{2}\frac{c_{\rm s}V_{\rm
h}}{\pi G\Sigma R},
\label{Q}
\end{equation}
where $R$ is the cylindrical radial coordinate, $\Sigma$ is the 
surface mass density, $c_{\rm s}$ is the sound speed, 
$\kappa=\sqrt{2}V_{\rm h}/R$ is the epicyclic frequency, and $V_{\rm h}$ 
is the circular velocity of the disc (mostly determined by the gravitational 
potential of the dark matter). It is also assumed that at the 
relevant radii ($\approx 10^2-10^3$ pc) the rotation curve is well 
described by a flat $V_{\rm h}$ profile.  We consider here the 
earliest generations of gas discs, which are of pristine composition 
with no metals and therefore can cool only via hydrogen. In thermal 
equilibrium, if the formation of molecular hydrogen is suppressed, 
these discs are expected to be nearly isothermal at a temperature of a 
few thousand Kelvin (here we take $T_{\rm gas}\approx 5000$K, LN06). 
However, molecular hydrogen if present can cool these 
discs further down to temperatures of a few hundred Kelvin. The 
stability parameter has a critical value $Q_{c}$ of the order of 
unity, below which the disc is unstable leading to the potential 
formation of a seed black hole. The actual value of $Q_{\rm c}$ 
essentially determines how stable the disc is, with lower $Q_{\rm c}$'s 
implying more stable discs. It is well known since Toomre (1964) 
proposed this stability criterion, that for an infinitesimally thin 
disc to be stable to fragmentation, $Q_{\rm c}=1$ for axisymmetric 
disturbances. The exact value of $Q_{\rm c}$ under more realistic 
conditions is not well determined. Finite thickness effects tend to 
stabilize the disc (reducing $Q_{\rm c}$), while on the other hand 
non-axisymmetric perturbations are in reality more unstable (enhancing 
$Q_{\rm c}$). Global, three-dimensional simulations of Keplerian discs 
(Lodato \& Rice 2004, 2005) have shown that such discs settle down in a
quasi-equilibrium configuration with $Q$ remarkably close to unity,
implying that the critical value $Q_{\rm c}\approx 1$. In this treatment, 
$Q_{\rm c}$ is taken to be a free parameter and results are computed
for a range of values. 

If the disc becomes unstable it develops non-axisymmetric spiral
structures, which leads to an effective redistribution of angular
momentum, thus feeding a growing seed black hole in the center.  This
process stops when the amount of mass transported to the center is 
enough to make the disc marginally stable. This can
be computed easily from the stability criterion in eqn.~(\ref{Q}) and
from the disc properties, determined from the dark matter halo mass
and angular momentum (Mo, Mao \& White 1998). The mass accumulated in the center of the halo (which
provides an upper limit to the SMBH seed mass) is given by:
\begin{equation}
M_{\rm BH}= m_{\rm d}M_{\rm halo}\left[1-\sqrt{\frac{8\lambda}{m_{\rm d}Q_{\rm c}}\left(\frac{j_{\rm d}}{m_{\rm d}}\right)\left(\frac{T_{\rm gas}}{T_{\rm vir}}\right)^{1/2}}\right] 
\label{mbh}
\end{equation}
for 
\begin{equation}
\lambda<\lambda_{\rm max}=m_{\rm d}Q_{\rm c}/8(m_{\rm d}/j_{\rm d}) (T_{\rm
  vir}/T_{\rm gas})^{1/2}
\label{lambdamax} 
\end{equation}
and $M_{\rm BH}=0$ otherwise. Here $\lambda_{\rm max}$ is the maximum
halo spin parameter for which the disc is gravitationally unstable,
$m_d$ is the gas fraction that participates in the infall and $Q_{\rm
c}$ is the Toomre parameter. The efficiency of SMBH formation is
strongly dependent on the Toomre parameter $Q_{\rm c}$, which sets the
frequency of formation, and consequently the number density of SMBH
seeds. The efficiency of the seed assembly process ceases at large halo
masses, where the disc undergoes fragmentation instead. This occurs
when the virial temperature exceeds a critical value $T_{\rm max}$,
given by:
\begin{equation}
\frac{T_{\rm max}}{T_{\rm gas}}=\left(\frac{4\alpha_{\rm c}}{m_{\rm
d}}\frac{1}{1+M_{\rm BH}/m_{\rm d}M_{\rm halo}}\right)^{2/3},
\label{frag}
\end{equation}
where $\alpha_{\rm c}\approx 0.06$ is a dimensionless parameter measuring the
critical gravitational torque above which the disc fragments.
The remaining relevant parameters are assumed to have typical values: $m_{\rm d}=j_{\rm d}=0.05$,
$\alpha_{\rm c}=0.06$ for the $Q_{\rm c}=2$ case.
The gas has a temperature $T_{\rm gas}=5000$K.

The process described above provides a means to transport matter from
a typical scale of a few hundred parsecs down to radii of a few AU. If
the halo-disc system already possesses a massive black hole seed from
a previous generation, then this gas can provide a large fuel
reservoir for its further growth. Note that the typical accretion
rates implied by the above model are of the order of
$0.01M_{\odot}/$yr, and are therefore sub-Eddington for seeds with
masses of the order of $10^5M_{\msun}$ or so. If, on the other hand no
black hole seed is present, then this large gas inflow can form a seed
anew. The ultimate fate of the gas in this case at the smallest scales
is more uncertain. One possibility, if the accretion rate is
sufficiently large, has been described in detail by Begelman et al. (2006). The 
infalling material likely forms a quasi-star,
the core of which collapses and forms a BH, while the quasi-star keeps
accreting and growing in mass at a rate which would be super-Eddington
for the central BH.  Alternatively, the gas might form a super-massive
star, which would eventually collapse and form a black hole (Shapiro \& Shibata 2002).  There 
are no quantitative estimates of how much
mass would ultimately end up collapsing into the hole. Thus, the black
hole seed mass estimates based on eqn.~(\ref{mbh}) should be considered 
as upper limits.

To give an idea of the efficiency of BH seed formation at high $z$
within the present model, we plot in Fig.~2, the probability of forming a 
BH (of any mass) at $z=18$, as a function of halo mass, for three illustrative 
models. It can be seen that typically up to 10\% of the halos in the right mass range can form a
central seed BH, the percentage rising to a maximum of $\approx$25\%
for the high efficiency Model C (highly unstable discs), and dropping
to a value of $\approx 4$\% for the high stability and therefore low
efficiency case (Model A). To summarize, every dark matter halo is characterized by its mass $M$
(or virial temperature $T_{\rm vir}$) and by its spin parameter
$\lambda$.  If $\lambda<\lambda_{\rm max}$ (see eqn.~\ref{lambdamax}) and $T_{\rm
vir}<T_{\rm max}$ (eqn.~\ref{frag}), then a seed SMBH forms in the center. 
Hence SMBHs form (i) only in halos within a given range of virial
temperatures, and hence, halo masses, and (ii) only within a narrow range of 
spin parameters, as shown in Fig.~3. High values of the spin
parameter, leading most likely to disk-dominated galaxies, are strongly disfavored as
seed formation sites in this model, and in models that rely on global dynamical 
instabilities.

\begin{figure}   
  \includegraphics[width=14cm]{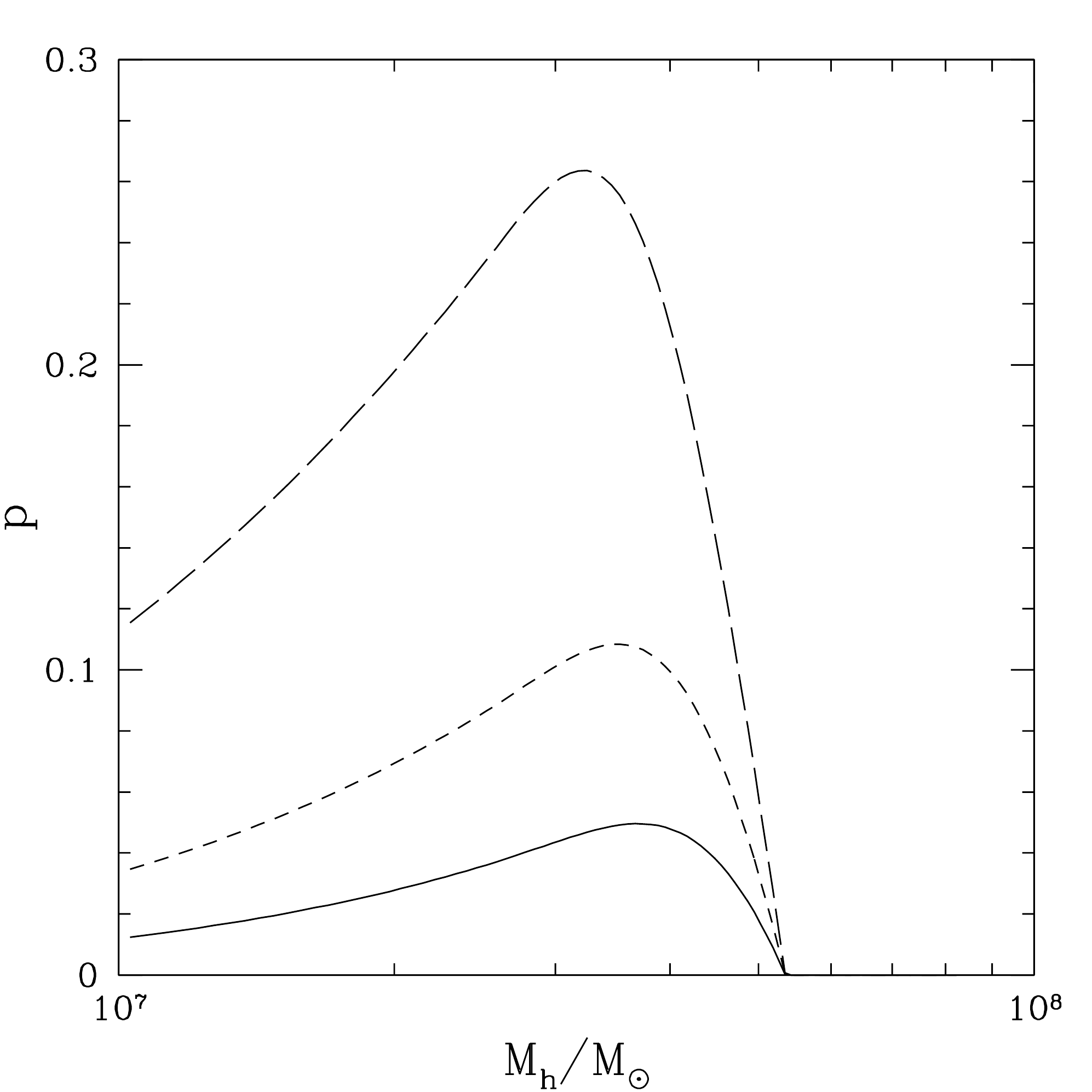}
  \caption{The probability of hosting a BH seed of any mass at $z=18$
    as a function of dark matter halo mass. The three curves refer to $Q_{\rm c}=1.5$
    (low efficiency case Model A, solid line), $Q_{\rm c}=2$ (intermediate efficiency case Model B,
    short-dashed line) and $Q_{\rm c}=3$ (high efficiency case Model C, long-dashed line).}
   \label{fig2}
\end{figure}

\begin{figure}
\includegraphics[width=14cm]{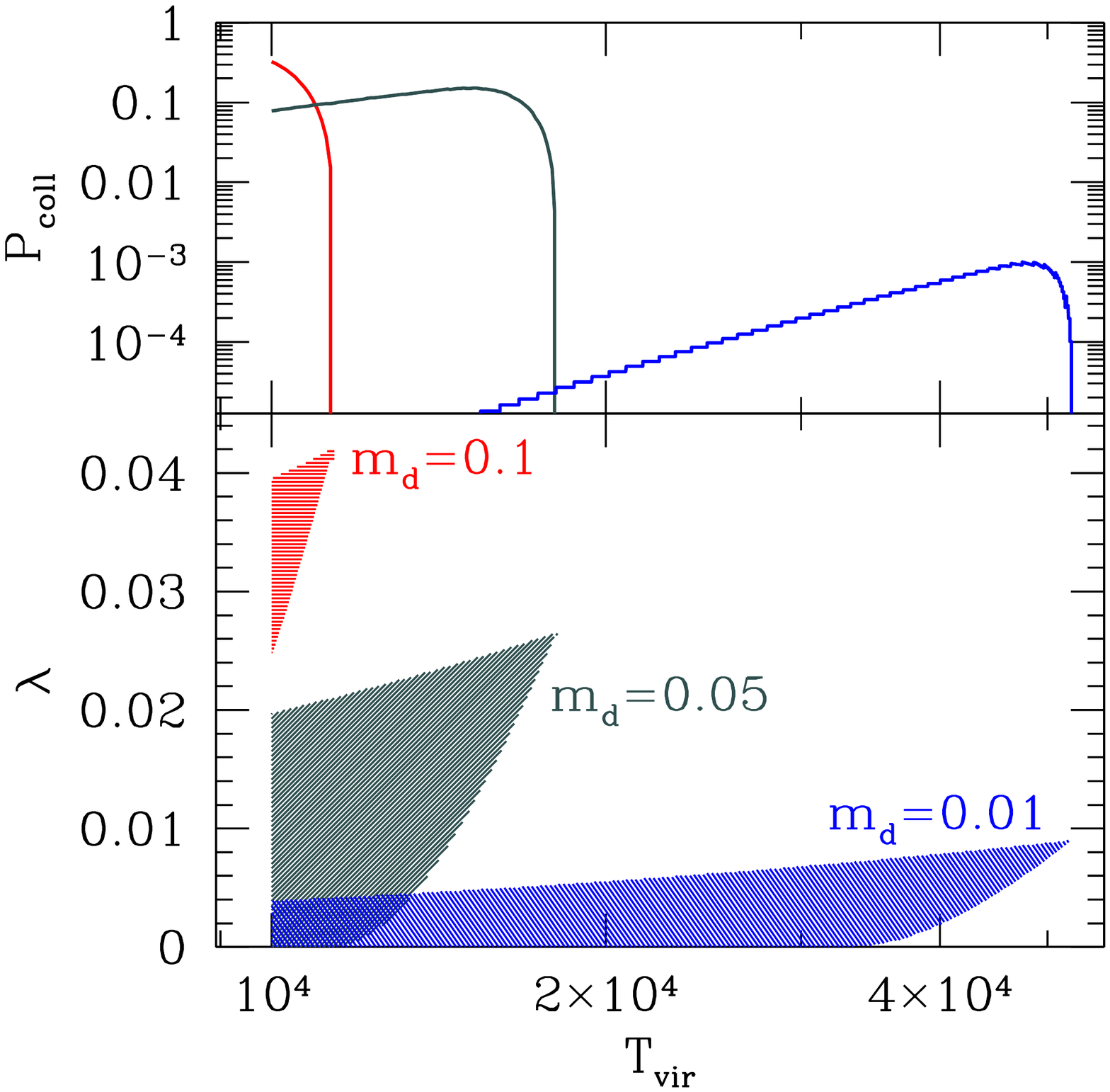}
\caption{Parameter space (virial temperature, spin parameter) for
  SMBH formation. Halos with $T_{vir}>10^4$ K
  at $z=15$are picked to participate in the infall ($m_{\rm d}$). 
  The shaded areas in the bottom panel show the range of virial temperatures and spin parameters 
  where discs are Toomre unstable and the joint conditions, $\lambda<\lambda_{\rm max}$
  (eqn.~\ref{lambdamax}) and $T_{\rm vir}<T_{\rm max}$ (eqn.~(\ref{frag}), showing the minimum spin 
  parameter, $\lambda_{\rm min}$ value below which the disc is globally prone to fragmantation) are fulfilled. 
  The top panel shows the probability of SMBH formation and is obtained, by integrating the lognormal distribution 
  of spin parameters between $\lambda_{\rm min}$ and  $\lambda_{\rm max}$.}
\label{fig3}
\end{figure}

\section{The evolution of seed black holes}

We follow the evolution of the MBH population resulting from the seed
formation process delineated above in a $\Lambda$CDM Universe. Our
approach is similar to the one described in Volonteri, Haardt \&
Madau (2003). We simulate the merger history of present-day halos with
masses in the range $10^{11}<M<10^{15}\,\msun$ starting from $z=20$,
via a Monte Carlo algorithm based on the extended Press-Schechter
formalism. Every halo entering the merger tree is assigned a spin parameter 
drawn from the lognormal $P(\lambda)$ distribution of simulated LCDM halos. 
Recent work on the fate of halo spins during mergers in cosmological simulations has led to
conflicting results: Vitvitska et al.  (2002) suggest that the spin
parameter of a halo increases after a major merger, and the angular
momentum decreases after a long series of minor mergers; D'Onghia \&
Navarro (2007) find instead no significant correlation between spin
and merger history. Given the unsettled nature of this matter,  we simply assume 
that the spin parameter of a halo is not modified by its merger history. When a halo enters 
the merger tree, we assign seed MBHs by determining
if the halo meets all the requirements described in Section 2 for the
formation of a central mass concentration. As we do not
self-consistently trace the metal enrichment of the intergalactic
medium, we consider here a sharp transition threshold, and assume that
the MBH formation scenario suggested by Lodato \& Natarajan ceases at
$z\approx 15$.  At $z>15$, therefore, whenever a new halo appears in the merger tree 
(because its mass is larger than the mass resolution), or a pre-existing 
halo modifies its mass by a merger, we evaluate if the gaseous component 
meets the conditions for efficient transport of angular momentum to create a
large inflow of gas which can either form a MBH seed, or feed one if
already present. 

The efficiency of MBH formation is strongly dependent on the critical value of 
the Toomre parameter $Q_{\rm c}$. We investigate the
influence of this parameter in the determination of the global
evolution of the MBH population. Fig.~4 shows the number
density of seeds formed in three different models with varying efficiency, 
with $Q_{\rm c}=1.5$ (low efficiency Model A), $Q_{\rm c}=2$ (intermediate 
efficiency Model B), and $Q_{\rm c}=3$ (high efficiency Model C). The solid 
histograms show the total mass function of seeds formed by $z=15$ when this
 formation channel ceases, while the dashed histograms refer to seeds formed at a
specific redshift slice at $z=18$.  The number of seeds changes by
about one order of magnitude from the least efficient to the most
efficient model, consistent with the probabilities shown in Fig.~2.

\begin{figure}   
   \includegraphics[width=14cm]{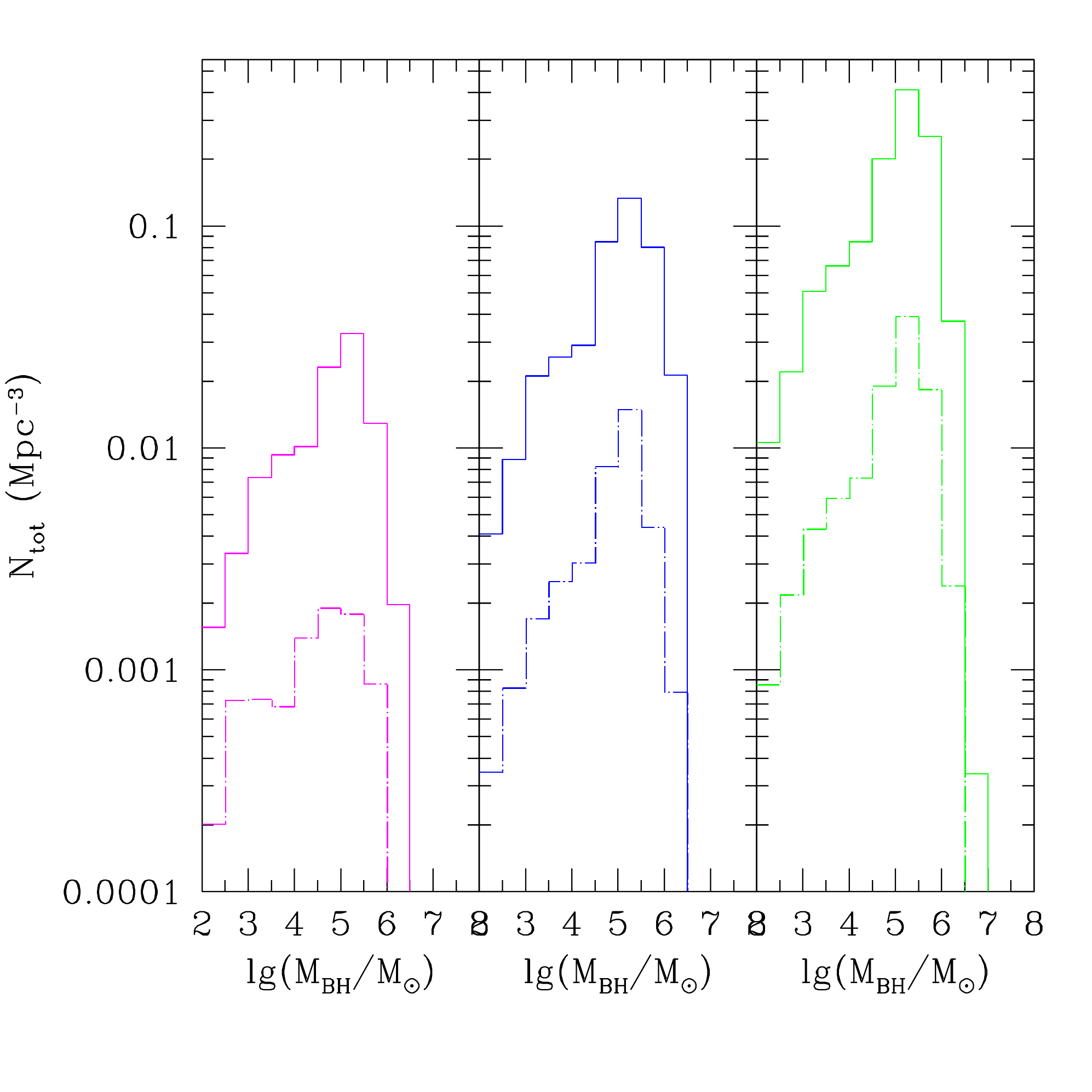} 
   \caption{Mass function of MBH seeds in the three Q-models of that
differ in seed formation efficiency. Left panel: $Q_{\rm c}=1.5$ (the
least efficient Model A), middle panel: $Q_{\rm c}=2$ (intermediate
efficiency Model B), right panel: $Q_{\rm c}=3$ (highly efficient
Model C). Seeds form at $z>15$ and this channel ceases at $z =
15$. The solid histograms show the total mass function of seeds formed
by $z=15$, while the dashed histograms refer to seeds formed at a
specific redshift, $z=18$.}
   \label{fig4}
\end{figure}

We assume that, after seed formation ceases, the $z<15$ population of
MBHs evolves according to a ``merger driven scenario", as described in Volonteri (2006). 
We assume that during major mergers MBHs accrete
gas mass that scales with the fifth power of the circular velocity
(or equivalently the velocity dispersion $\sigma_c$) of the host halo. We thus set 
the final mass of the MBH at the end of the
accretion episode to 90\% of the mass predicted by the $M_{\rm BH}-\sigma_c$
correlation, assuming that the scaling does not evolve with
redshift.  Mergers of the BHs themselves contribute to the
mass increase of the remaining 10\%.

We briefly outline the merger scenario calculation here. The merger rate of halos can be estimated 
using eqn.~(1) of Fakhouri et al. (2010), where a simple fitting formula is derived from large LCDM simulations.
The merger rate per unit redshift and mass ratio ($\xi$) at fixed halo mass is given by:
\beq
\frac{dN_m}{d\xi dz}(M_h) = A\left(\frac{M_h}{10^{12} M_0}\right)^{\alpha}\xi^{\beta}\exp\left[\left(\frac{\xi}{\tilde{\xi}}\right)^{\gamma}\right](1+z)^{\eta}.
\label{mjm}
\eeq
with A=0.0104, $\alpha=0.133$, $\beta=-1.995$, $\gamma=0.263$, $\eta=0.0993$ and $\tilde{\xi}=9.72\times10^{-3}$. We can integrate the merger rate between $z=0$ and say, $z=3$,
and mass ratio $\xi >0.3$ (major mergers). This gives the number of major mergers a halo of a given mass experiences between $z=0$ and $z=3$. 
Halo mass can be translated into virial circular velocity:
\begin{equation}
 V_{\rm c}= 142 {\rm km/s} \left[\frac{M_{\rm h}}{10^{12} \ M_{\odot} }\right]^{1/3} 
\left[\frac {\Omm}{\Ommz}\ \frac{\Delta_{\rm c}} {18\pi^2}\right]^{1/6} 
(1+z)^{1/2},  
\end{equation}
where $\Delta_{\rm c}$ is the over--density at virialization relative to the critical density. 
For a WMAP5 cosmology we adopt here the  fitting
formula  $\Delta_{\rm c}=18\pi^2+82 d-39 d^2$ (Bryan \& Norman 1998), 
where $d\equiv \Ommz-1$ is evaluated at the collapse redshift, so
that $ \Ommz={\Omm (1+z)^3}/({\Omm (1+z)^3+\Oml+\Omk (1+z)^2})$.  It is well known that the 
major merger rate is an increasing function of halo mass or circular velocity, in fact we 
find that the expected number of mergers between $z=0$ and $z=3$ with mass ratio $\xi >0.3$ is
$\simeq$~0.4 for $M_h=10^8 \msun$, $\simeq$~0.5 for $M_h=10^9 \msun$, $\simeq$~0.7 for $M_h=10^{10} \msun$, $\simeq$~1.0 for $M_h=10^{11} \msun$,
$\simeq$~1.4 for $M_h=10^{12} \msun$, $\simeq$~1.8 for $M_h=10^{13} \msun$.  

In order to calculate the accreting black hole luminosity function and
to follow the black hole mass growth during each accretion event, we
also need to calculate the rate at which the mass, as estimated above,
is accreted. This is assumed to scale with the Eddington rate for the
MBH, and is based on the results of merger simulations, which 
heuristically track accretion onto a central MBH (Di Matteo et al. 2005; Hopkins et al. 2005; 
Sijacki, Springel, Di Matteo \& Hernquist  2007).  The time spent by a given simulated AGN 
at a given bolometric luminosity \footnote{We convert accretion rate 
into luminosity assuming that the radiative efficiency equals the 
binding energy per unit mass of a particle in the last stable circular 
orbit. We associate the location of the last stable circular orbit to 
the spin of the MBHs, by self-consistently tracking the evolution of 
black hole spins throughout our calculations (Volonteri 2005). We set 20\% as the maximum value of the 
radiative efficiency, corresponding to a spin slightly below the 
theoretical limit for thin disc accretion(Thorne 1974). } per logarithmic interval is approximated by 
Hopkins et al. (2005) as follows:
\begin{equation}
\label{eq:dtdlogL }
\frac{{d}t}{{d}L}=|\alpha|t_Q\, L^{-1}\, \left(\frac{L}{10^9L_\odot}\right)^\alpha,
\end{equation}
where $t_Q\simeq10^9$ yr, and $\alpha=-0.95+0.32\log(L_{\rm
peak}/10^{12} L_\odot)$. Here $L_{\rm peak}$ is the luminosity of the
AGN at the peak of its activity.  Hopkins et al. (2006) show that
approximating $L_{\rm peak}$ with the Eddington luminosity of the MBH
at its final mass (i.e., when it sits on the $M_{\rm BH}-\sigma_c$
relation) compared to computing the peak luminosity with eqn.~(6)
above gives the same result and in fact, the difference between these
2 cases is negligible. Volonteri et al. (2006) derive the following
simple differential equation to express the instantaneous accretion
rate ($f_{\rm Edd}$, in units of the Eddington rate) for a MBH of mass
$M_{\rm BH}$ in a galaxy with velocity dispersion $\sigma_c$:
\begin{equation}
\label{eq:doteddratio }
\frac{{d}f_{\rm Edd}(t)}{{d}t}=\frac{ f_{\rm Edd}^{1-\alpha}(t) }{|\alpha|
  t_Q} \left(\frac{\epsilon \dot{M}_{\rm Edd} c^2}{10^9L_\odot}\right)^{-\alpha},
\end{equation}
where $t$ is the time elapsed from the beginning of the accretion event.
Solving this equation provides us the instantaneous Eddington ratio for a
given MBH at a specific time, and therefore we can self-consistently grow
the MBH mass. We set the Eddington ratio $f_{\rm Edd}=10^{-3}$ at
$t=0$. This same type of accretion is assumed to occur, at $z>15$,
following a major merger in which a MBH is not fed by disc
instabilities. 

\begin{figure}   
\includegraphics[width=14cm]{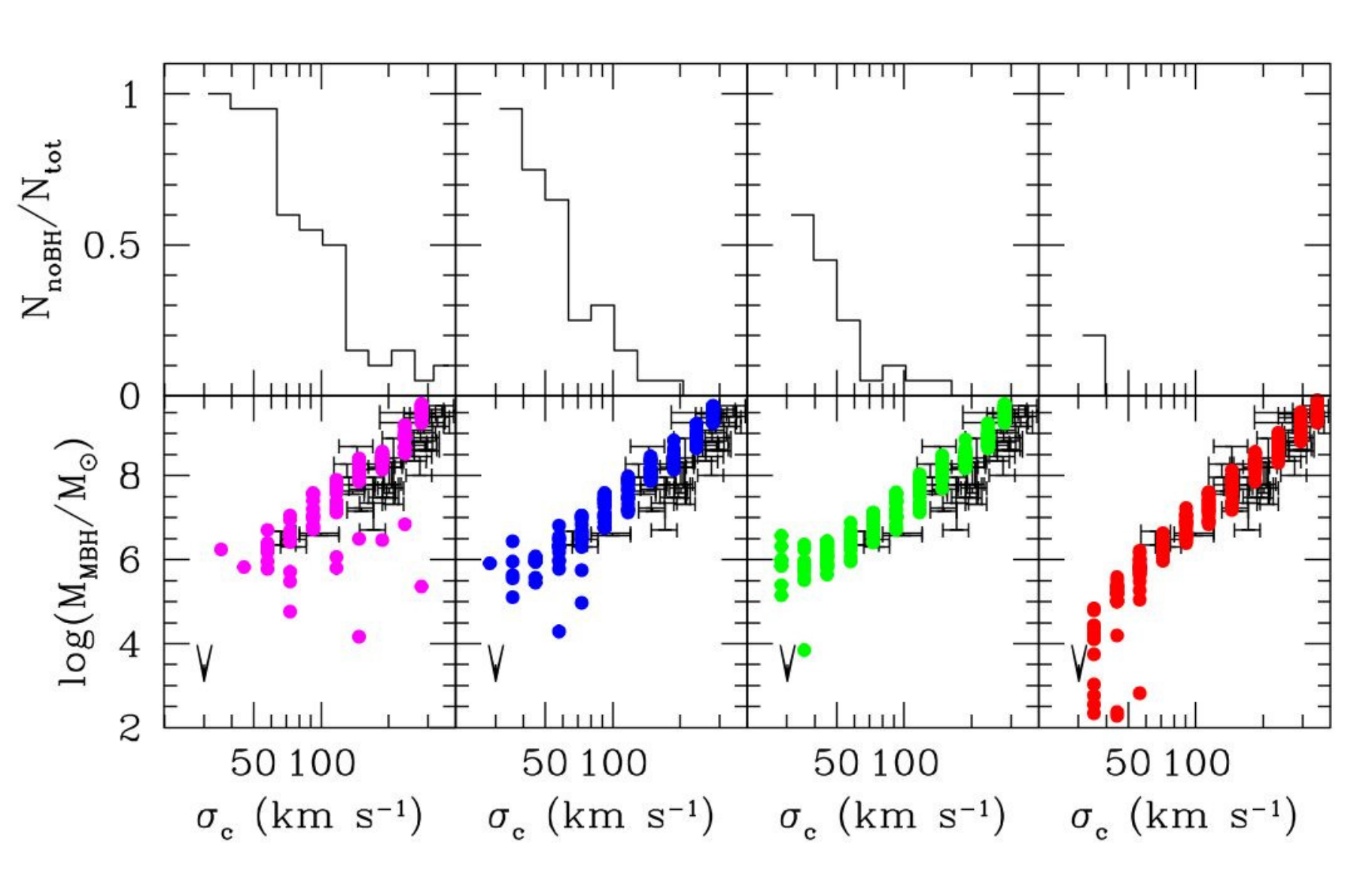} 
\caption{ The $M_{\rm bh}-$velocity dispersion ($\sigma_c$)
     relation at $z=0$. Every circle represents the central MBH in a
     halo of given $\sigma_c$.  Observational data are marked by their
     quoted errorbars, both in $\sigma_c$, and in $M_{\rm bh}$
     (Tremaine et al. 2002).  Left to right panels: $Q_{\rm c}=1.5$,
     $Q_{\rm c}=2$, $Q_{\rm c}=3$, Population-III remnant seeds.  {\it
     Top panels:} fraction of galaxies at a given velocity dispersion
     which {\bf do not} host a central MBH.}
\label{fig5}
\end{figure}

We present the results of tracking the assembly history of MBHs in 2
classes of galaxies, (i) a dark matter halo with mass
$2\times10^{12}\,M_{\odot}$ that hosts a MW type galaxy and 
(ii) a more massive dark matter halo, $4\times10^{13}\,M_{\odot}$, that hosts
a massive early type (ET) galaxy (see Fig.~6). The progenitors of the MBHs in each of these host
halos are tracked and plotted as measured at a given epoch. We
analyze 20 realizations for each halo, to account for cosmic variance.
Examples of growth histories are shown in
Fig.~6, while statistical $\msigma$ relations are shown for the two seed models.

As outlined earlier, in propagating the seeds it is assumed that
accretion episodes and therefore growth spurts are triggered only by
major mergers. We find that in a merger-driven scenario for MBH growth
the most biased galaxies at every epoch host the most massive MBHs
that are most likely already sitting on the $M_{\rm BH} -\sigma$
relation.  Lower mass MBHs (below $10^6\,\msun$) are instead off the
relation at $z = 4$ and even at $z = 2$. These baseline results are
{\it independent of the seeding mechanism}.  
In the `heavy seeds' scenario, most of the MBH seeds start out {\it
well above} the $z=0$ $\msigma$, that is, they are `overmassive'
compared to the local relation.  Seeds form only in halos within a
narrow range of velocity dispersion ($\sigma \simeq 15\,{\rm
km\,s}^{-1} $, see equations~1 and 3, and Fig.~1).  The MBH mass
corresponding to $\sigma \simeq 15\,{\rm km\,s}^{-1}$, according to
the local $M_{\rm BH} -\sigma$ relation, would be $\sim 3\times 10^3
\msun$.  The initial BH seed mass function instead peaks at $10^5\msun$ (Lodato \&
Natarajan 2007).  As time elapses, all halos are bound to grow in
mass by mergers.  The lowest mass halos, though, experience mostly
minor mergers, that do not trigger accretion episodes, and hence do
not grow the MBH. The evolution of these systems can be described by
a shift towards the right of the $\msigma$ relation: $\sigma$
increases, but $M_{\rm BH}$ stays roughly constant. Such systems are
clearly seen at $z=1$ in Fig.~6, with $M_{\rm
BH}\sim10^5\msun$ and $\sigma<100\,{\rm km\,s}^{-1}$.  Effectively,
for the lowest mass halos growth of the stellar component of the galaxy and the central MBH
are not coeval but rather sequential.

In the case of Population-III seeds there is initially no
correlation between seed mass and halo mass or velocity
dispersion. Here we have assumed that the seeds form in the mass range
$125<M_{\rm BH} <1000\,\msun$.  The initial $\msigma$ relation would
therefore appear as a horizontal line at $\sim 200\msun$. In this case MBHs migrate
onto the $\msigma$ always from {\it below}, as seeds are initially
`undermassive' compared to the local relation. Underfed survivors of
the seed epoch shift towards the right of the $\msigma$ relation and
lie in the lower left corner of Fig.~\ref{fig2}, with $M_{\rm
BH}\sim10^2-10^3\msun$ and $\sigma<100\,{\rm km\,s}^{-1}$.

There appears to be a distinct difference between the journey of MBH
seeds onto the $\msigma$ relation predicted by the two seeding models
considered here. The Population-III seeds start life `undermassive'
lying initially below the local $\msigma$ and they transit up to the
relation by essentially growing the MBH without significantly altering
$\sigma$. In contrast, the massive seeds start off above the local
$\msigma$ relation, and migrate onto it by initially growing $\sigma$,
after which further major mergers trigger accretion episodes and
therefore growth spurts for the MBHs.  When MBHs are more massive than
expected compared to the $\msigma$ relation, accretion is terminated very rapidly in 
our scheme (physically, we expect feedback to be responsible for shutting
down accretion, see, e.g., Silk \& Rees 1998, Fabian 2002; Natarajan \& Treister 2009).

\begin{figure}   
  \includegraphics[width=14cm]{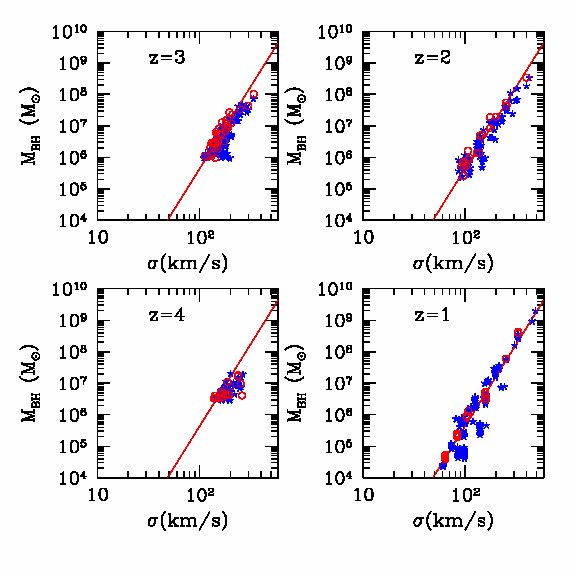}
  \caption{The $\msigma$ relation for active MBHs at different
  redshift slices in the ET progenitors. These MBHs would be observed
  as AGNs.  We imposed a flux threshold, 10$^{-16}$ erg s$^{-1}$ cm$^{-2}$ (bolometric). 
  Stars: Massive seeds based on the model by Lodato \&
  Natarajan (2006), with $Q_c=2$. Circles: seeds based on Population
  III star remnant models (Volonteri et al. 2003).  The figure shows
  both central and satellite MBHs (satellite holes are shown at the
  $\sigma$ of the host halo). The sample comprises all the progenitors
  of 20 $z=0$ halos.}
   \label{fig6}
\end{figure}

It appears that major mergers that trigger accretion episodes are what
set up the relation initially at high redshift (Treister et al. 2010). Our conclusions in
this regard are in agreement with those reached by alternative
arguments, for instance see Peng (2007) and Robertson et
al. (2006). Recent results from X-ray stacking analysis of the $z ~ 7 - 8$ 
dropouts is also consistent with the existence of self-regulation even
at these early times (Treister et al. 2011).  Biased peaks in the halo mass distribution, which are the
sites for the formation of the largest galaxies, host the earliest
massive MBHs that fall on the relation. Hence, the $\msigma$ 
relation is established first for MBHs hosted in the largest halos
present at any time. MBHs in small galaxies lag behind, as their hosts
are subject to little or no major merger activity. In many cases the
MBHs remain at the original seed mass for billions of years (e.g., see
Figs.~6 \& 7, the $z=1$ panel).  We find that these conclusions
hold irrespective of our initial seeding mechanism and the relation
tightens considerably from $z = 4$ to $z = 1$, especially for MBHs
hosted in halos with $\sigma>100$ $\rm{km\, s}^{-1}$.  We find that
if black hole seeds are massive, $\sim 10^5\,M_{\odot}$, the low-mass
end of the $\msigma$ flattens at low masses towards an asymptotic
value, creating a characteristic `plume'. This `plume' consists of
ungrown seeds, that merely continue to track the peak of the seed mass
function at $M_{\rm BH}\sim10^5\,\msun$ down to late times. For the
Population-III seed case, since the initial seed mass is very small,
the plume of MBHs with $M_{\rm BH}\sim10^5-10^6\,\msun$ in halos with
$\sigma\sim 40-50\,{\rm km\, s^{-1}}$ disappears.

\section{Results}

The repercussions of different initial efficiencies for seed formation
for the overall evolution of the MBH population stretch from
high-redshift to the local Universe.
Detection of gravitational waves from seeds merging at the redshift of
formation (Sesana 2007) is probably one of the best ways to
discriminate among formation mechanisms. On the other hand, the
imprint of different formation scenarios can also be sought in
observations at lower redshifts. The various seed formation scenarios
have distinct consequences for the properties of the MBH population at
$z=0$ as we demonstrate below. 

Given the accuracy of current observational constraints seeding models
cannot be differentiated. Discrimination between the models appears
predominantly at the low mass end of the present day black hole mass
function which is not observationally well constrained. However, all
our models predict that low surface brightness, bulgeless galaxies
with large discs are least likely to be sites for the formation of
massive seed black holes at high redshifts. The efficiency of seed
formation at high redshifts has a direct influence on the black hole
occupation fraction in galaxies at $z=0$. This effect is more
pronounced for low mass galaxies.  This is the key discriminant
between the models studied here and the Population-III remnant seed
model as we show in detail below.  We find that there exists a population of low mass galaxies
that do not host nuclear black holes. Our prediction of the shape of
the $M_{\rm bh} - \sigma$ relation at the low mass end is in agreement
with the recent observational determination from the census of low
mass galaxies locally (Greene \& Ho 2007; Kormendy \& Bender 2011).

\subsection{Predictions at high redshift}

\subsubsection{The luminosity function of accreting black holes}

Turning to the global properties of the MBH population, as suggested
by Yu \& Tremaine (2002) the mass growth of the MBH population at
$z<3$ is dominated by the mass accreted during the bright epoch of
quasars, thus washing out most of the imprint of initial conditions.
This is evident when we compute the luminosity function of AGN.
Clearly the detailed shape of the predicted luminosity function
depends most strongly on the accretion prescription used. With our
assumption that the gas mass accreted during each merger episode is
proportional to $V_c^5$, we find that distinguishing between the
various seed models is difficult. As shown in Fig.~\ref{fig4}, all 3
models reproduce the bright end of the observed bolometric LF
(Hopkins et al. 2007) at higher redshifts (marked as the solid curve in
all the panels), and predict a fairly steep faint end that is as yet
undetected. All models fare less well at low redshift, shown in
particular at $z = 0.5$. This could be due to the fact that we have
used a single accretion prescription to model growth through
epochs. On the other hand, the decline in the available gas budget at
low redshifts (since the bulk of the gas has been consumed by this
epoch by star formation activity) likely changes the radiative
efficiency of these systems. Besides, observations suggest a sharp
decline in the number of actively accreting black holes at low
redshifts across wave-lengths, produced most probably due to changes
in the accretion flow as a result of change in the geometry of the
nuclear regions of galaxies. In fact, all 3 of our models
under-predict the slope at the faint end. There are three other effects
that could cause this flattening of the LF at the faint end at low
redshift for our models: (i) not having taken into account the fate of
on-going merging and the fate of satellite galaxies; (ii) the number
of realizations generated and tracked is insufficient for statistics,
as evidenced by the systematically larger error-bars and (iii) more
importantly, it is unclear if merger-driven accretion is indeed the
trigger of BH fueling in the low redshift universe. We note that the 3
massive seed models and Population-III seed model cannot be
discriminated by the LF at high redshifts. Models B and C are also in
agreement viz-a-viz the predicted BH mass function at $z = 6$, even
assuming a very high radiative efficiency (up to 20\%), while Model A might
need less severe assumptions, in particular for BH masses larger than
$10^7\,\msun$.

\begin{figure}
   \includegraphics[width=14cm]{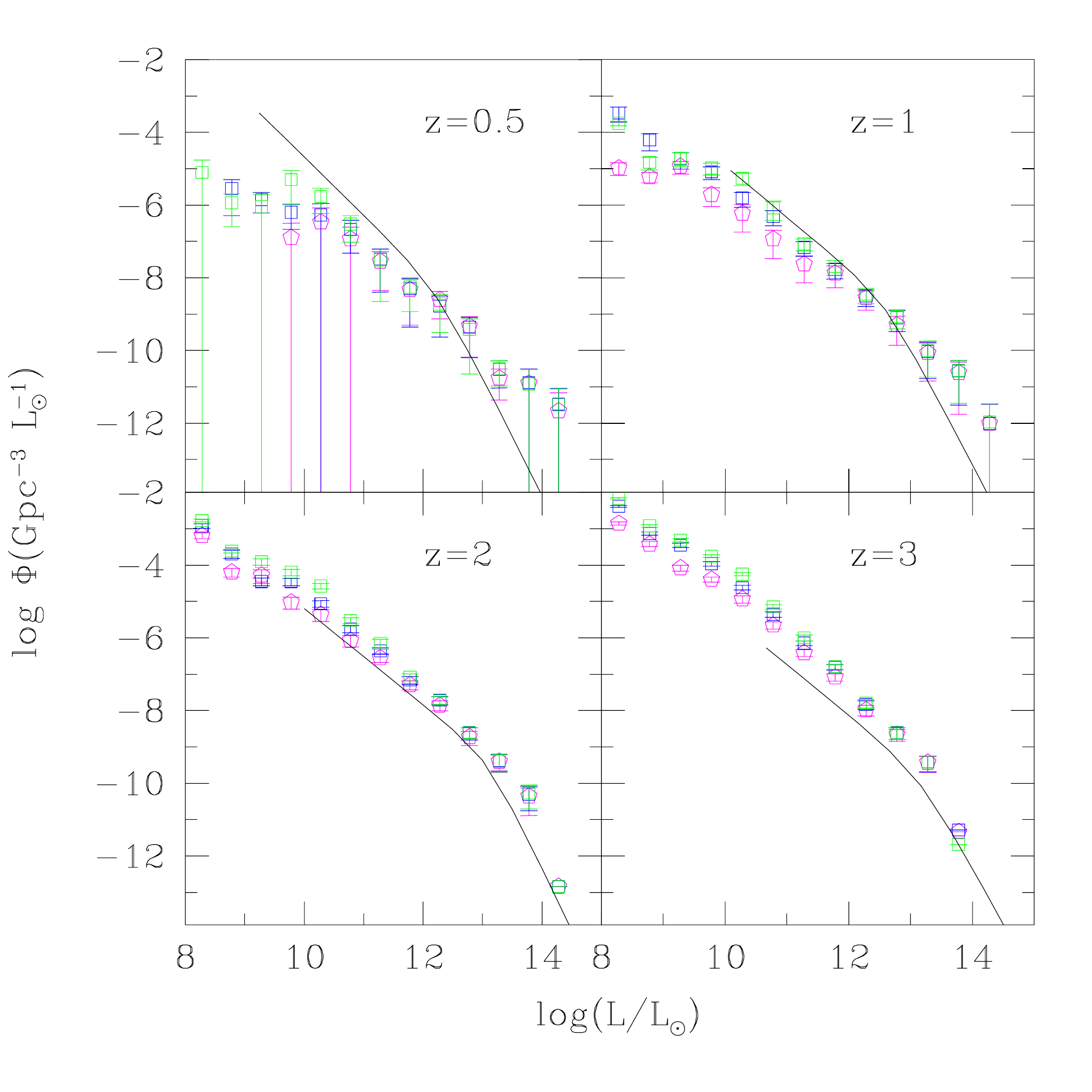} 
   \caption{Predicted bolometric luminosity functions at different
redshifts with observational data over-plotted. All 3 models match the observed bright end of the LF at
high redshifts and predict a steep slope at the faint end down to $z =
1$. The 3 models are not really distinguishable with the LF. However
at low redshifts, for instance at $z = 0.5$, all 3 models are
significantly flatter at both high and low luminosities and do not
adequately match the current data. As discussed in the text, the LF is
strongly determined by the accretion prescription and what we see here
is simply a reflection of that fact.}
   \label{fig7}
\end{figure}

\subsection{Low redshift predictions}

\subsubsection{Supermassive black holes in dwarf galaxies}

Obviously, a higher density of MBH seeds implies a more ubiquitous population 
of MBHs at later times, which can produce observational signatures in statistical
samples. More subtly, the formation of seeds in a $\Lambda$CDM
scenario follows the cosmological bias. As a consequence, the
progenitors of massive galaxies have a higher probability of hosting MBH 
seeds (cfr. Madau \& Rees 2001). In the case of low-bias systems, such as isolated dwarf 
galaxies, very few of the high-$z$ progenitors have the deep potential wells needed
for gas retention and cooling, a prerequisite for MBH formation. 
In the lowest efficiency massive seed Model A, for example, a galaxy needs of order 25 massive progenitors (mass
above $\sim10^7\msun$) to ensure a high probability of seeding within
the merger tree. In the massive seed Model C, instead, the requirement drops to 4
massive progenitors, increasing the probability of MBH formation in
lower bias halos.

The signature of the efficiency of the formation of MBH seeds will
consequently be stronger in isolated dwarf galaxies. Fig.~5
(bottom panel) shows a comparison between the observed $M_{\rm
BH}-\sigma$ relation and the one predicted by our models (shown with
circles), and in particular, from left to right, the three models
based on the LN06 and Lodato \& Natarajan (2007) seed masses with $Q_{\rm c}=1.5$, 2 and
3, and a fourth model based on lower-mass Population-III star
seeds. The upper panel of Fig.~5 shows the fraction
of galaxies that {\bf do not} host any massive black holes for different
velocity dispersion bins. This shows that the fraction of galaxies
without a MBH increases with decreasing halo masses at $z = 0$. 
A larger fraction of low mass halos are devoid of central black holes for
lower seed formation efficiencies. Note that this is one of the key
discriminants between our models and those seeded with Population-III
remnants. As shown in Fig.~3, there are practically no galaxies without
central BHs for the Population-III seeds. We can therefore make quantitative predictions for 
the local occupation fraction of MBHs. Our massive seeding Model A predicts that 
below $\sigma_c\approx 60\,{\rm kms}^{-1}$ the
probability of a galaxy hosting a MBH is negligible. With increasing
MBH formation efficiencies, the minimum mass for a galaxy that hosts a
MBH decreases, and it drops below our simulation limits for Model
C. On the other hand, models based on lower mass Population-III star
remnant seeds, predict that massive black holes might be present even
in low mass galaxies. These predictions appear to be consistent with recent
observations of low mass galaxies (Kormendy \& Bender 2011).

Although there are degeneracies in our modeling (e.g., between the
minimum redshift for BH formation and instability
criterion), the BH occupation fraction, and the masses of the BHs in
dwarf galaxies are the key diagnostics.  An additional caveat worth mentioning is the
possibility that a galaxy is devoid of a central MBH because of
dynamical ejections (due to either the gravitational recoil or
three-body scattering). The signatures of such dynamical interactions
should be more prominent in dwarf galaxies, but ejected MBHs would
leave observational signatures on their hosts (G{\"u}ltekin et al. in
prep.). On top of that, Schnittman (2007) and Volonteri, Lodato \& Natarajan (2007)
agree in considering the recoil a minor correction to the overall
distribution of the MBH population at low redshift (cfr. Fig.~ 4 in
Volonteri 2007). Additionally, as MBH seed formation requires halos with low angular
momentum (low spin parameter), we envisage that low surface
brightness, bulge-less galaxies with high spin parameters (i.e. large
discs) are systems where MBH seed formation is less
probable. Furthermore, bulgeless galaxies are believed to preferentially have
quieter merger histories and are unlikely to have experienced 
major mergers, which could have brought in a MBH from a companion
galaxy. This prediction of the massive seeding models appears to be
in consonance with recent observational studies by Kormendy \& Bender (2011).

 \subsubsection{Seed signatures for the detection of gravitational waves}

An alternative to AGN observations in electromagnetic bands is the
detection of MBHs via gravitational radiation, that would be
detectable by proposed future missions like {\it LISA}. The merger rate of MBHs in our models, and the detectability of 
binaries has been discussed in Sesana et al. (2007), where the impact of different `seed' 
formation scenarios was taken into account. Since the focus here is on high-redshift objects, we assume that merging is driven by 
dynamical friction, which has been shown to efficiently  drive the MBHs 
in the central regions of the newly formed  galaxy when the mass ratio of the satellite halo to the 
main halo is sufficiently  large, $>\,1:10$ 
and galaxies are gas--rich (Callegari et al. 2008).  The available simulations 
(Armitage \& Natarajan 2002; Escala et al. 2004; Dotti et al. 2006; Mayer et al. 2006) 
show that the binary can shrink to about parsec or slightly subparsec scale by 
dynamical friction against gas.  We refer the reader to  Sesana et al. (2007) and Sesana et al. (2005)
for a detailed discussion of how we model the gravitational wave emission and the 
expected event rate. Detection of gravitational radiation provides accurate measurements of 
the mass of the components of MBH binaries prior to  merger, and the mass of the single 
merger remnant. Additionally, the mass of `single' MBHs can be determined by the inspiral 
of an extreme or intermediate mass-ratio compact object (EMRI/IMRI, Miller 2005).

When we track the merging population, we find that MBH-MBH mergers
also preferentially sample the region of space
where MBHs lie on the $\msigma$ relation. This is once again a
consequence of halo bias.  Both channels for seed formation that we investigate here 
require deep potential wells for gas retention and
cooling as a prerequisite for MBH formation. Halos where massive
seeds can form are typically 3.5--4 $\sigma$ peaks of the density
fluctuation field at $z>15$, (the host halos in the direct collapse
model are slightly more biased than in the Population-III remnant
case). MBH seeding is therefore infrequent, MBHs are rare and as a
consequence MBH-MBH mergers are events that typically involve only the
most biased halos at any time. 

In typical mergers we find that the higher mass black hole in the binary
tends to sit on or near the expected $\msigma$ relation for
the host (which corresponds to the newly formed galaxy after the merger).  
The mass of the secondary generally provides clues to the
dynamics of the merger, rather than to the $\msigma$ relation, since
at the time of the merger any information that we can gather on the host
(via electromagnetic observations) will not provide details on the two original galaxies.  
For instance the mass ratio of the merging MBHs encodes how efficiently minor
mergers can deliver MBHs to the centre of a galaxy in order to form a
bound binary.

\subsubsection{Hidden black holes}

One key finding is the prediction of the existence of a large
population of hidden (as in undetectable as AGN or as merging BHs via
gravitational radiation) MBHs at all redshifts. There are two main contributors to the population of
hidden MBHs: MBHs in the \textit{nuclei} of low-mass galaxies
($\sigma\sim 20-50 {\rm km\, s^{-1}}$), and satellite/wandering MBHs.
`Hidden' nuclear MBHs have not experienced appreciable growth in mass
and formed in low mass halos with quiet merging histories. A
potential observational signature of the massive seed scenario is the
existence of a `plume' of overmassive MBHs in the \textit{nuclei} of
halos with $\sigma\sim 20-50\, {\rm km\, s^{-1}}$.  The only way to
detect MBHs in the plume would be as IMRI/EMRI (intermediate or
extreme mass ratio inspiral) events, or via measurement of stellar
velocity dispersions and modelling as in the local universe (for
example Magorrian et al. 1998).  Approaching $z = 0$, the under-fed
part of this population likely merges into more massive galaxies.
 
Satellite and wandering MBHs would instead be off-centre systems, orbiting
in the potential of comparatively massive hosts.  Semantically we
distinguish here between MBHs that are infalling into a galaxy for the
very first time, following a galaxy merger (satellite MBHs) and those
that are merely displaced from the center due to gravitational recoil
(wandering MBHs).  Some of the {\it satellite} MBHs will merge with
the central MBH in the primary galaxy, and such merging does not
significantly alter the position of the already massive primary hole
which sits on the $M_{\rm BH} - \sigma$ relation to start with.

The MBH population in our series of simulations of the massive ET halo
is shown in Fig.~6, for $z=1$. Here we dissect the MBH
population into its components.  Satellite/wandering MBHs are found
{\it below} the $\msigma$ correlation as expected (shown as open
circles, at the $\sigma$ of the host halo). Luminous AGNs are
preferentially found on the $\msigma$ relation (squares). We note the
existence of a sub-population of satellite AGNs, that is, satellite
MBHs which are actively accreting. For every pair of coalescing MBHs
(triangles), one typically sits on the $\msigma$ relation, while the
companion tends to be less massive, hence, when they merge, the
remnant finds itself in the right spot on the $\msigma$ relation
(solid circles).

\subsubsection{Comoving mass density of black holes}
 
Since during the quasar epoch MBHs increase their mass by a large factor,
signatures of the seed formation mechanisms are likely more evident at
{\it earlier epochs}. We compare in Fig.~8 the integrated
comoving mass density in MBHs to the expectations from So{\l}tan-type
arguments, assuming that quasars
are powered by radiatively efficient flows (for details, see Yu \& Tremaine 2002; 
Elvis, Risaliti \&  Zamorini 2002; Marconi et al. 2004). While during and after the quasar
epoch the mass densities in models A, B, and C differ by less than a
factor of 2, at $z>3$ the differences are more pronounced.

A very efficient seed MBH formation scenario can lead to a very large
BH density at high redshifts. For instance, in the highest efficiency
Model C with $Q_{\rm c}=3$, the integrated MBH density at $z=10$ is
already $\sim 25\%$ of the density at $z=0$. The plateau at $z>6$ is
due to our choice of scaling the accreted mass with the $z=0$ $M_{\rm
bh}-\sigma$ relation. In our models we let MBHs
accrete mass that scales with the fifth power of the circular
velocity of the halo, the accreted mass is a small fraction of the MBH
mass (see the discussion in Marulli et al. 2006), and the overall
growth remains small, as long as the mass of the seed is larger than
the accreted mass, which, for our assumed scaling, happens whenever  the
mass of the halo is below a few times $10^{10}\msun$. The comoving
mass density, an integral constraint, is reasonably well determined
out to $z = 3$ but is poorly known at higher redshifts. All models
appear to be satisfactory and consistent with current observational
limits (shown as the shaded area).

\begin{figure}   
   \includegraphics[width=14cm]{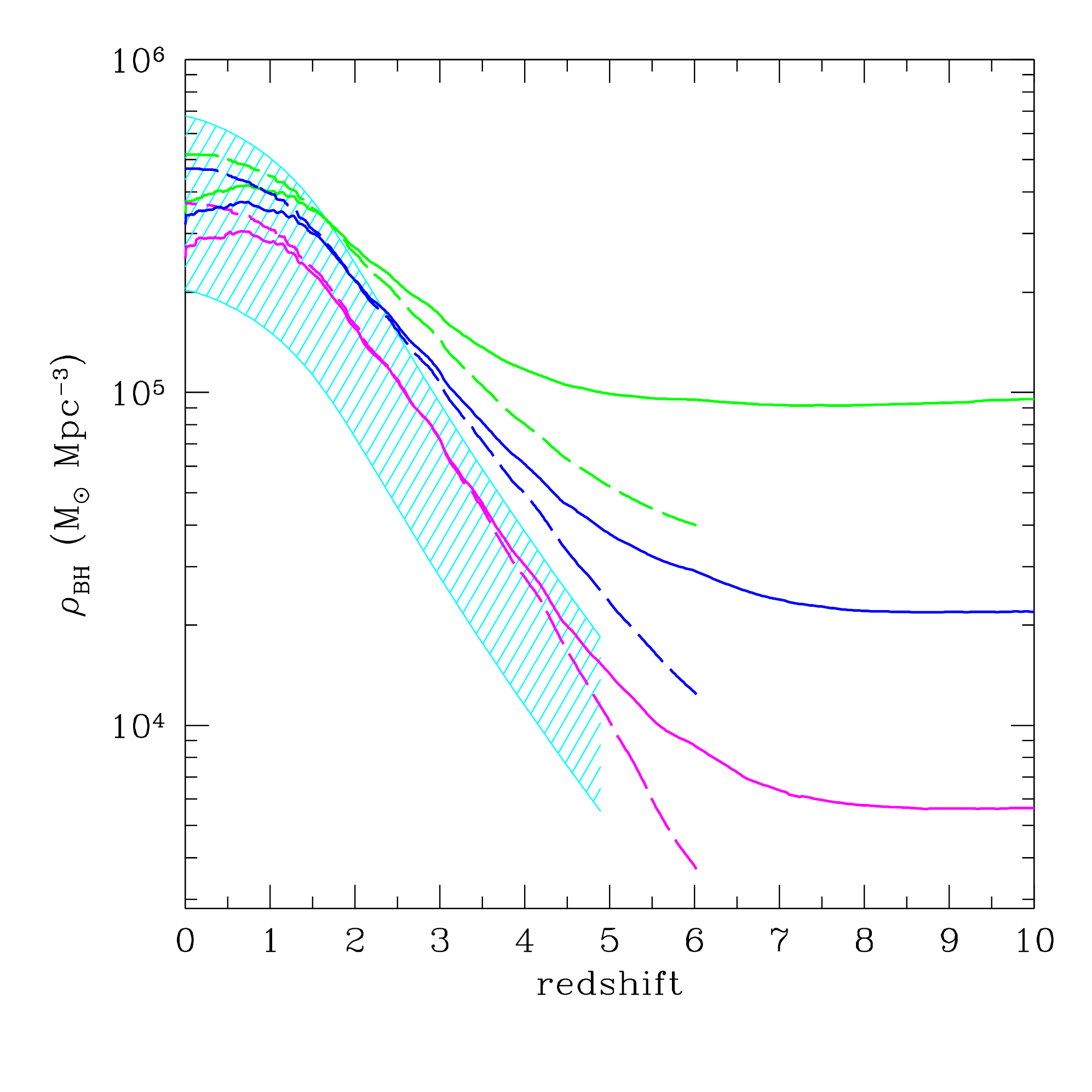} 
   \caption{Integrated black hole mass density as a function of
     redshift. Solid lines: total mass density locked into nuclear
     black holes.  Dashed lines: integrated mass density {\rm
     accreted} by black holes.  Models based on BH remnants of
     Population-III stars (lowest curve), $Q_{\rm c}=1.5$ (middle
     lower curve), $Q_{\rm c}=2$ (middle upper curve), $Q_{\rm c}=3$
     (upper curve).  Shaded area: constraints from So{\l}tan-type
     arguments, where we have varied the radiative efficiency from a
     lower limit of 6\% (applicable to Schwarzschild MBHs, upper
     envelope of the shaded area), to about 20\%. All 3 massive seed formation models are in comfortable
     agreement with the mass density obtained from integrating the
     optical luminosity functions of quasars.}
   \label{fig8}
\end{figure}

\subsubsection{Black hole mass function at $z=0$}

One of the key diagnostics is the comparison of the measured and
predicted BH mass function at $z = 0$ for our 3 models. In
Fig.~9, we show (from left to right, respectively) the mass
function predicted by models A, B, C and Population-III remnant seeds
compared to that obtained from measurements.  The histograms show the
mass function obtained with our models (where the upper histogram
includes all the black holes while the lower one only includes black
holes found in central galaxies of halos in the merger-tree
approach). The two lines are two different estimates of the observed
black hole mass function. In the upper one, the measured velocity
dispersion function for nearby late and early-type galaxies from the
SDSS survey (Bernardi et al. 2003; Sheth et al. 2003) has been convolved with the
measured $M_{\rm BH} - \sigma$ relation. We note here that the scatter
in the $M_{\rm bh} - \sigma$ relation is not explicitly included in
this treatment, however the inclusion of the scatter is likely to
preferentially affect the high mass end of the BHMF, which provides
stronger constraints on the accretion histories rather than the seed
masses. It has been argued by Tundo et al. (2007); Bernardi et al. (2007) 
and Lauer et al. (2007) 
that the BH mass function differs if the bulge mass is used instead of the 
velocity dispersion in relating the BH mass to the host galaxy. 
Since our models do not trace the formation and growth of stellar bulges 
in detail, we are restricted to using the velocity dispersion in our analysis.

The lower dashed curve is an alternate theoretical estimate of the BH mass 
function derived using the Press-Schechter formalism from  Jenkins et al. (2001) 
in conjunction with the observed $M_{\rm BH} -\sigma$ relation. Selecting only the
central galaxies of halos in the merger-tree approach adopted here
(lower histograms) is shown to be equivalent to this analytical
estimate, and this is clearly borne out in the plot.
When we include black holes in satellite galaxies (upper histograms,
cfr. the discussion in Volonteri, Haardt \& Madau 2003) the predicted
mass function moves towards the estimate based on SDSS galaxies. The
higher efficiency models clearly produce more BHs. At higher
redshifts, for instance at $z = 6$, the mass functions of active MBHs
predicted by all models are in very good agreements in particular for
BH masses larger than $10^6\,\msun$, as it is the growth by accretion
that dominates the evolution of the population. At the highest mass
end ($>10^9\,\msun$) Model A lags behind models B and C, although we
stress once again that our assumptions for the accretion process are
very conservative.

The {\it relative} differences between Models A, B, and C at the
low-mass end of the mass function, however, are genuinely related to
the MBH seeding mechanism (see also Figs.~2 \& 4). In Model A, simply, fewer galaxies host a MBH, hence
reducing the overall number density of black holes. Although our
simplified treatment does not allow robust quantitative predictions,
the presence of a "bump" at $z = 0$ in the MBH mass function at the
characteristic mass that marks the peak of the seed mass function
(cfr. Fig.~4) is a sign of highly efficient formation of
massive seeds (i.e., much larger mass with respect, for instance,
Population-III remnants). The higher the efficiency of seed formation,
the more pronounced is the bump (note that the bump is most prominent
for Model C). Since current measurements of MBH masses extend barely
down to $M_{\rm bh}\sim 10^6 \msun$, this feature cannot be
observationally tested with present data but future campaigns, with
the Giant Magellan Telescope or JWST, are likely to extend the mass
function measurements to much lower black hole masses.

\begin{figure}   
   \includegraphics[width=14cm]{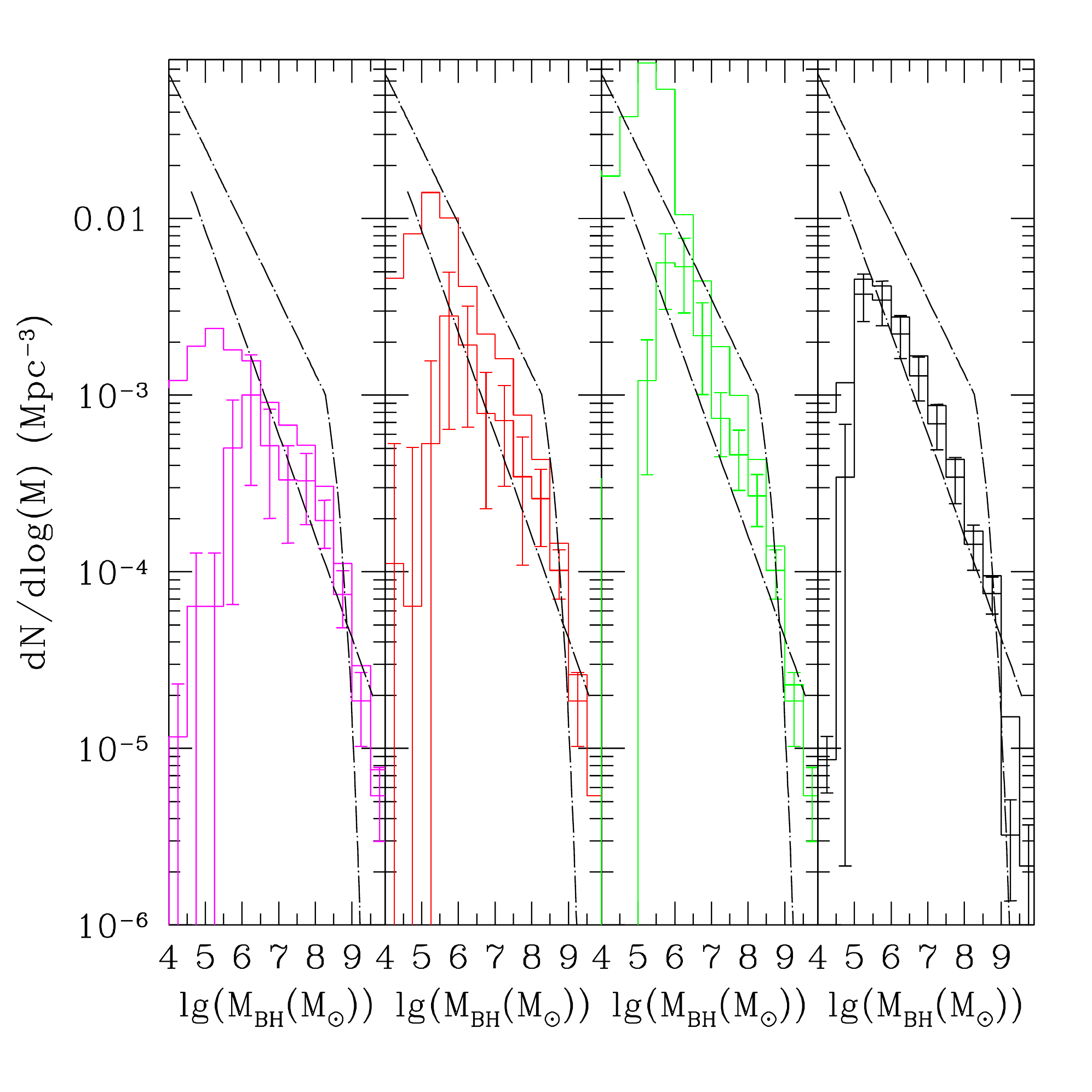} 
   \caption{Mass function of black holes at z=0. Histograms represent
     the results of our models, including central galaxies only (lower
     histograms with errorbars), or including satellites in groups and
     clusters (upper histograms). Left panel: $Q_{\rm c}=1.5$, mid-left
     panel: $Q_{\rm c}=2$, mid-right panel: $Q_{\rm c}=3$, right panel:
     models based on BH remnants of Population-III stars.  
     Upper dashed line: mass function derived from combining the velocity
     dispersion function of Sloan galaxies (Sheth et al. 2003, where
     we have included the late-type galaxies extrapolation), and BH
     mass-velocity dispersion correlation (e.g., Tremaine et
     al. 2002). Lower dashed line: mass function derived using the
     Press-Schechter formalism from Jenkins et al. (2001) in
     conjunction with the $M_{\rm BH} -\sigma$ relation (Ferrarese
     2002).}
   \label{fig9}
\end{figure}

\section{Conclusions}

Ih this review, we outline massive black hole seed formation models 
and focus on the consequences on their assembly history as a function
of cosmic epoch. While the errors on mass determinations of local black holes
are large at the present time, definite trends with host galaxy
properties are observed. The tightest correlation appears to be between
the BH mass and the velocity dispersion of the host spheroid. Starting
with the ab-initio black hole seed mass function computed in the
context of direct formation of central objects from the collapse of
pre-galactic discs in high redshift halos, we follow the assembly
history to late times using a Monte-Carlo merger tree approach. Key to
our calculation of the evolution and build-up of mass is the
prescription that we adopt for determining the precise mass gain
during a merger. Motivated by the phenomenological observation of
$M_{\rm BH} \propto V_{\rm c}^5$, we assume that this proportionality
carries over to the gas mass accreted in each step. With these
prescriptions, a range of predictions can be made for the mass
function of black holes at high and low $z$, and the integrated mass
density of black holes, all of which are observationally
determined. We evolve 3 models, designated Model A, B and C which
correspond to increasing efficiencies respectively for the formation
of seeds at high redshift. These models are compared to one in which
the seeds are remnants of Population-III stars. 

It is important to
note here that one major uncertainty prevents us from making more
concrete predictions: the unknown metal enrichment history of
the Universe. Key to the implementation of our models is the choice of
redshift at which massive seed formation is quenched. The direct
seed formation channel described here ceases to operate once the Universe
has been enriched by metals that have been synthesized after the 
first generation of stars have gone supernova. Once metals are available
in the Inter-Galactic Medium gas cooling is much more efficient and hydrogen
in either atomic or molecular form is no longer the key player. In this 
work, we have assumed this transition redshift to be $z = 15$. The efficiency 
of MBH formation and the transition redshift are somewhat degenerate (e.g., 
a model with $Q=1.5$ and enrichment redshift $z=12$ is halfway between 
Model A and Model B); if other constraints on this redshift were available we 
could considerably tighten our predictions. 
 
Below we list our predictions and compare how they fare with 
current observations. The models investigated here clearly differ notably
in predictions at the low mass end of the black hole mass function. 
With future observational sensitivity in this domain, these models 
can be distinguished.

\begin{enumerate}

\item{Occupation fraction at $z = 0$: Our model for the formation of relatively high-mass 
black hole seeds in high-$z$ halos has direct influence on the black hole occupation
fraction in galaxies at $z=0$. All our models predict that low surface brightness, bulge-less
  galaxies with high spin parameters (i.e. large discs) are systems
  where MBH formation is least probable.  We find that a significant fraction of low-mass
galaxies might not host a nuclear black hole. This is in very good
agreement with the shape of the $M_{\rm bh} - \sigma$ relation
determined recently from an observational census (an HST ACS survey)
of low mass galaxies in the Virgo cluster reported by Ferrarese et
al. (2006). While current data in the low mass
  regime is scant (Barth 2004; Greene \& Ho 2007; Kormendy \& Bender 2010), future instruments and surveys 
  are likely to probe this region of parameter space with significantly higher
  sensitivity.
}

\item{High mass end of the local SMBH mass function: While the models studied here (with different black hole seed formation
  efficiency) are distinguishable at the low mass end of the BH mass
  function, at the high mass end the effect of initial seeds
  appears to be sub-dominant. These models cannot be easily distinguished 
  by observations at $z \sim 3$.}
  
  \end{enumerate}
  
  One of the key caveats of our picture is that it is unclear if the
  differences produced by different seed models on observables at 
  $z = 0$ might be compensated or masked by BH fueling modes at earlier
  epochs. There could be other channels for BH growth that dominate at
  low redshifts like minor mergers, dynamical instabilities, accretion of
  molecular clouds, tidal disruption of stars. The decreased importance of the
  merger driven scenario is patent from observations of low-redshift AGN, 
  which are for the large majority hosted by undisturbed galaxies (e.g. Pierce 
  et al. 2007 and references therein) in 
  low-density environments. However, the 
  feasibility and efficiency of some alternative channels are still to 
  be proven (for example the efficiency of feeding from large scale 
  instabilities see discussion in King \& Pringle 2007; Shlosman 1989; Goodman 2003; Collin 1999). 
  In any event, while these additional channels for BH {\it growth} can modify the 
  detailed shape of the mass function of MBHs, or of the luminosity function of 
  quasars, they will not create new MBHs.  The occupation fraction 
  of MBHs (see Fig.~ 5) is therefore largely {\it independent} of the 
  accretion mechanism and a true signature of the formation process.
    
 To date, most theoretical models for the evolution of MBHs in galaxies
do not include {\it how} MBHs form. Our work offers a first analysis of
the observational signatures of massive black hole formation
mechanisms in the low redshift universe, complementary to the
investigation by Sesana et al. (2007), where the focus was on
detection of seeds at the very early times where they form, via
gravitational waves emitted during MBH mergers. We focus here on
possible dynamical signatures that forming massive black hole seeds
imprint on the local Universe. Obviously, the signatures of
seed formation mechanisms will be far more clear if considered jointly
with the evolution of the spheroids that they host. The mass, and
especially the frequency, of the forming MBH seeds is a necessary
input when investigating how the feedback from accretion onto MBHs
influences the host galaxy, and is generally introduced in numerical
models using extremely simplified, {\it ad hoc} prescriptions (e.g.,
Springel et al. 2005; Di Matteo, Springel \& Hernquist 2005; Hopkins et al. 2006;
Croton et al. 2005; Cattaneo et al 2006; Bower et al. 2006).
Adopting more detailed models for black hole seed formation, as outlined 
here, can in principle strongly affect such results. Incorporating  sensible 
assumptions for the masses, and
frequency of MBH seeds in models of galaxy formation is necessary if
we want to understand the symbiotic growth of MBHs and their hosts. Much work remains 
to be done in order to obtain a deeper understanding of the accretion 
history of BHs over cosmic time.



\begin{theacknowledgments}
 PN would like to acknowledge her collaborators Marta Volonteri and
Giuseppe Lodato with whom most of this work was 
done. She would also to thank the John Simon Guggenheim Foundation
for support via a Guggenheim fellowship and the Institute for Theory
and Computation at Harvard University for hosting her.
 \end{theacknowledgments}







\IfFileExists{\jobname.bbl}{}
 {\typeout{}
  \typeout{******************************************}
  \typeout{** Please run "bibtex \jobname" to optain}
  \typeout{** the bibliography and then re-run LaTeX}
  \typeout{** twice to fix the references!}
  \typeout{******************************************}
  \typeout{}
 }

\end{document}